\documentclass[12pt]{article}
\usepackage{epsf}

\renewcommand{\thefootnote}{\#\arabic{footnote}}
\begin{document}

\newcommand{\gtrsim}{ \mathop{}_{\textstyle \sim}^{\textstyle >} }
\newcommand{\lesssim}{ \mathop{}_{\textstyle \sim}^{\textstyle <} }

\renewcommand{\thefootnote}{\fnsymbol{footnote}}
\setcounter{footnote}{0}
\begin{titlepage}

\def\thefootnote{\fnsymbol{footnote}}

\begin{center}

\hfill TU-603\\
\hfill hep-ph/0010197\\
\hfill October, 2000\\

\vskip .5in

{\Large \bf
  Constraints on Natural Inflation\\
  from Cosmic Microwave Background
}

\vskip .45in

{\large
  Takeo Moroi and Tomo Takahashi
}

\vskip .45in

{\em
  Department of Physics,  Tohoku University, Sendai 980-8578, Japan
}

\end{center}

\vskip .4in

\begin{abstract}

  We study constraints on the natural inflation model from the cosmic
  microwave background radiation (CMBR).  Inflaton $\phi$ for the
  natural inflation has a potential of the form
  $V=\Lambda^4[1-\cos(\phi/\sqrt{2}f_\phi)]$, which is parametrized by
  two parameters $f_\phi$ and $\Lambda$.  Various cosmological
  quantities, like the primordial curvature perturbation and the CMBR
  anisotropy, are determined as functions of these two parameters.
  Using recent observations of the CMBR anisotropy by BOOMERANG and
  MAXIMA (as well as those from COBE), constraints on the parameters
  $f_\phi$ and $\Lambda$ are derived.  The model with $f_\phi$ lower
  than $8.5\times 10^{18}$ ($5.4\times 10^{18}$, $4.5\times 10^{18}$)
  GeV predicts a power spectrum with index $n_{\rm S}$ smaller than
  0.95 (0.9, 0.85) which suppresses the CMBR anisotropy for smaller
  angular scale.  With such a small $n_{\rm S}$, height of the second
  acoustic peak can become significantly lower than the case of the
  scale-invariant Harrison-Zeldovich spectrum.

\end{abstract}
\end{titlepage}

\renewcommand{\thepage}{\arabic{page}}
\setcounter{page}{1}
\renewcommand{\thefootnote}{\#\arabic{footnote}}
\setcounter{footnote}{0}

Inflation \cite{PRD23-347} plays a very important role in modern
cosmology.  It provides a natural solution to the horizon and flatness
problems.  In addition, quantum fluctuation during the inflation
becomes the origin of the density perturbation of the universe.  Such
a fluctuation also generates temperature perturbation in the cosmic
microwave background radiation (CMBR) which is being measured very
accurately by the satellite and balloon experiments.

In the case of the ``slow-roll inflation''
\cite{PLB108-389,PRL48-1220}, where the inflation is due to the energy
density of a slowly-rolling scalar field, the scalar field, which is
called ``inflaton,'' has to have a very flat potential.  It is,
however, difficult to guarantee the flatness of the potential because
radiative corrections to the scalar mass are in general quadratically
divergent.  Thus the most natural value of the scalar mass is as large
as the cutoff scale which is naturally the Planck scale.  With such a
large mass parameter, it is impossible to build a realistic model of
inflation.

One way to obtain a very flat potential is to consider the ``natural
inflation'' \cite{PRL65-3233,PRD47-426} where a (pseudo-)
Nambu-Goldstone (NG) field is used as the inflaton.\footnote{Another
way to guarantee the flatness is to introduce supersymmetry.  In the
supersymmetric models, quadratic divergences are cancelled out among
bosonic and fermionic loops.  In this paper, we do not consider such
a possibility.} If a global Abelian symmetry, which we call
$U(1)_X$, is spontaneously broken, massless scalar field shows up due
to the Nambu-Goldstone's theorem.  If the $U(1)_X$ symmetry is {\sl
  softly} broken, such a scalar field acquires a non-vanishing
potential, and the height of the potential is controlled by
soft-breaking parameters.  In this case, hierarchy between the scalar
mass and the Planck (or the cutoff) scale is naturally stabilized.

In the natural inflation, the inflaton $\phi$ has a particular form of
the potential: $V=\Lambda^4[1-\cos(\phi /\sqrt{2}f_\phi)]$.
Therefore, observable quantities are quite predictive and it is
interesting to study its consequences.  Indeed, the density
fluctuation from the natural inflation was already extensively
discussed in connections with the large scale structure and the COBE
observations \cite{PRD47-426}.  Recently, however, BOOMERANG
\cite{Nature404-955} and MAXIMA \cite{aph0005123} reported more
accurate observations of the CMBR anisotropy up to the multipole
$l\sim O(100)$.  With those new observations, we can have a better
constraint on the natural inflation scenario.  (For discussions on
other inflation models, see Refs.\ \cite{aph0007375,aph0008165}.)

In this letter, we consider constraints on the natural inflation model
from the CMBR. For an accurate estimation of the power spectrum, we
numerically follow the evolution of the inflaton field. The power
spectrum deviates from the scale-invariant Harrison-Zeldovich spectrum
when $f_\phi$ is relatively small, and the CMBR anisotropy from the
natural inflation can be significantly different from that from the
scale-invariant power spectrum.  Furthermore, taking the constraint on
the height of the second acoustic peak seriously, we may obtain an
upper bound on the scale $f_\phi$.

We consider a model where NG boson is used as the inflaton.  As an
example, let us start with a model with the following scalar potential
\begin{eqnarray}
V (X) = \lambda \left( |X|^2 - v^2 \right)^2
+ A M_*^{4-N} \left( X^N + {X^N}^\dagger \right),
\label{V(X)}
\end{eqnarray}
where $X$ is a complex scalar field, and $N$ is a positive integer.
We assume that smallness of model parameters should be protected by
some symmetry; thus, we take $\lambda\sim O(1)$ and $v$ of the order
of the reduced Planck scale $M_*\simeq 2.4\times 10^{18}\ {\rm GeV}$.
On the contrary, when $A=0$, there is an Abelian symmetry to rotate
the phase of $X$, which we call $U(1)_X$ symmetry.  Smallness of the
$A$ parameter is protected by $U(1)_X$.  In the following, we consider
the case with $A\ll 1$.

The $U(1)_X$ symmetry is broken by the vacuum expectation value of
$X$.  In this case, it is convenient to parameterize the $X$ field as
\begin{eqnarray}
X = \left( v + \frac{\sigma}{\sqrt{2}v} \right)
e^{i\phi / \sqrt{2}v},
\end{eqnarray}
where $\sigma$ and $\phi$ are real scalar fields.  The real component
$\sigma$ acquires a mass as large as $v$ and is irrelevant for our
discussion.  On the contrary, the imaginary component $\phi$ is the
(pseudo-) NG mode and is massless if $A=0$.  When $A\neq 0$, $\phi$
acquires a potential as\footnote{There may be other terms with higher
periodicity in general.  If the potential of $\phi$ is due to
$U(1)_X$ breaking spurion, however, the potential of $\phi$ is
expected to be of the form of Eq.\ (\ref{V_phi}).  For example, if
$A$ originates to a spurion with charge $N$, sub-leading term is
$\sim A^2\cos(\sqrt{2}\phi/f_\phi+\alpha)$, where $\alpha$ is an
unknown phase.  For realistic natural inflation, $A\ll 1$ and the
sub-leading contribution is negligible.}
\begin{eqnarray}
V(\phi ) = \Lambda^4
\left[ 1 - \cos\left(\frac{\phi}{\sqrt{2}f_\phi}\right) \right]
= 2\Lambda^4 
\sin^2\left(\frac{\phi}{2\sqrt{2}f_\phi}\right),
\label{V_phi}
\end{eqnarray}
where we added a constant to the potential for the vanishing
cosmological constant, and
\begin{eqnarray}
\Lambda^4 = \frac{2Av^N}{M_*^{N-4}},~~~
f_\phi = \frac{v}{N}.
\end{eqnarray}
As one can see, the height of the potential is controlled by the soft
breaking parameter $A$ and the potential can be very flat with a small
value of $A$.  Since the flatness of the potential is protected by the
(approximate) symmetry, quantum correction does not destabilize the
flatness.  The potential given in Eq.\ (\ref{V_phi}) is our starting
point.

$V(\phi)$ has a minimum at $\phi=0$ (mod $2\sqrt{2}\pi f_\phi$).
Expanding the potential around the minimum, we obtain $V
=\frac{1}{2}m_\phi^2\phi^2+O(\phi^4)$, where the mass of $\phi$ is
given by
\begin{eqnarray}
m_\phi = \frac{\Lambda^2}{\sqrt{2}f_\phi}.
\label{m_phi}
\end{eqnarray}
At this level, $\phi$ is stable.  If $\phi$ couples to other fields,
however, it may decay.  For example, $X$ may couple to a fermion
$\psi_Q$ with standard-model gauge quantum numbers with the
interaction ${\cal L}_{\rm int}=y_QX\bar{\psi}_Q\psi_Q+{\rm h.c.}$
(Here, $y_Q$ is a coupling constant.)  Once $X$ acquires the vacuum
expectation value, $\psi_Q$ becomes massive and the process
$\phi\rightarrow\bar{\psi}_Q\psi_Q$ is kinematically blocked.  At the
one-loop level, however, $\phi$ decays into the standard-model gauge
boson pairs.  The decay rate for this process is given by
\begin{eqnarray}
\Gamma_\phi = \sum_{i=1}^{3}
\frac{{\rm dim}(G_i)}{32\pi} 
\left( \frac{b^{(i)}_{\bar{\psi}\psi}\alpha_i}{4\pi N} \right)^2
\frac{m_\phi^3}{f_\phi^2},
\label{Gamma_phi}
\end{eqnarray}
where $\alpha_i$ is the coupling constant for the gauge group
$G_i=SU(3)_{\rm C}, SU(2)_{\rm L}$, and $U(1)_{\rm Y}$ while
$b^{(i)}_{\bar{\psi}\psi}$ the $\beta$-function coefficient of
$\bar{\psi}_Q+\psi_Q$.  Here, the sum is over the standard model gauge
groups.  In our calculation, we approximate the formula as
$\Gamma_\phi=10^{-5}m_\phi^3/f_\phi^2$.

We use the potential (\ref{V_phi}) and identify the $\phi$ field as
the inflaton.  We started with a specific example of the potential
(\ref{V(X)}).  However, notice that the following discussion is
independent of the structure of the underlying model; once the
potential (\ref{V_phi}) is given, the following results all hold.

If the $\phi$ field is displaced from the minimum in the early
universe, it rolls towards the minimum of the potential.  When
$t\lesssim\Gamma_\phi^{-1}$, the scalar field $\phi$ obeys the
following equation of motion:
\begin{eqnarray}
\ddot{\phi} + 3 H \dot{\phi} + V' = 0,
\label{eom_phi}
\end{eqnarray}
where the ``dot'' is the derivative with respect to time $t$ while
$V'\equiv dV/d\phi$.  Here, $H$ is the expansion rate $H=\dot{a}/a$
with $a$ being the scale factor.

When the displacement of $\phi$ from the minimum is larger than $\sim
M_*$, slow-roll conditions, $\epsilon\ll 1$ and $\eta\ll 1$, are
satisfied, where
\begin{eqnarray}
\epsilon = \frac{1}{2} \left( \frac{M_* V'}{V} \right)^2,~~~
\eta = \frac{M_*^2 V''}{V}.
\end{eqnarray}
Then, the energy density of the universe is dominated by the {\sl
  potential} energy of the inflaton field, and the universe is in the
de Sitter phase.  In this case, $\ddot{a}>0$ and the comoving scale
grows faster than the horizon.

When $\phi$ becomes less than the Planck scale, the slow-roll
conditions do not hold.  Then, $\ddot{a}$ becomes negative and the
inflation ends.  This happens when
\begin{eqnarray}
\dot{\phi}^2 = V.
\label{end_inf}
\end{eqnarray}
After this epoch, the scalar field starts to oscillate and the
amplitude of the oscillation decreases.  Eventually, the amplitude
becomes so small that the scalar potential is well-approximated as
$V\simeq \frac{1}{2}m_\phi^2\phi^2$.  Then the energy density of
$\phi$ is proportional to $a^{-3}$ (as far as the decay process is
neglected).  In this period, we can solve the Boltzmann equation for
the energy density of $\phi$ instead of following the motion of
$\phi$, since the change of the energy density $\rho_\phi$ during the
time scale $m_\phi^{-1}$ becomes very small:
\begin{eqnarray}
\dot{\rho}_\phi + 3 H \rho_\phi = -\Gamma_\phi \rho_\phi.
\label{drho_phi}
\end{eqnarray}
In addition, the radiation energy density $\rho_{\rm rad}$ obeys
\begin{eqnarray}
\dot{\rho}_{\rm rad} + 4 H \rho_{\rm rad} = \Gamma_\phi \rho_\phi.
\label{drho_rad}
\end{eqnarray}
When $t\sim\Gamma_\phi^{-1}$, the inflaton decays. Then the energy
density of $\phi$ is converted to that of the radiation and the
universe is reheated.  The reheating temperature is approximately
given by $T_{\rm R}\sim\sqrt{\Gamma_\phi M_*}$.

We follow the evolution of the inflaton field numerically.  Before
showing the numerical results, however, it is instructive to discuss
the qualitative behavior of the result with some approximation.  To
make the situation clear, we consider the case where the initial value
of the inflaton to be $0<\phi<\sqrt{2}\pi f_\phi$, although the
results do not depend on this assumption.  Solving Eq.\ 
(\ref{end_inf}) with the slow-roll approximation, we obtain the
amplitude at the end of the inflation as
\begin{eqnarray}
\phi_{\rm end} \simeq 2 \sqrt{2}f_\phi
\tan^{-1} \left( \frac{M_*}{\sqrt{6}f_\phi} \right).
\label{phi_end}
\end{eqnarray}
When $\phi>\phi_{\rm end}$, $\ddot{a}>0$ and the universe inflates.
The $e$-folding number during the inflation is estimated as
\begin{eqnarray}
N_e (\phi) \simeq \frac{2f_\phi^2}{M_*^2} \ln 
\left[ \frac{\cos^2(\phi_{\rm end}/2\sqrt{2}f_\phi)}
{\cos^2(\phi/2\sqrt{2}f_\phi)} \right].
\label{N_e}
\end{eqnarray}
The scale of the COBE observation ($k_{\rm COBE}\simeq 7.5a_0H_0$
\cite{PRep314-1} with $H_0\equiv 100h\ {\rm km/sec/Mpc}$ and $a_0$
being the present expansion rate and the scale factor, respectively)
exits the horizon when $N_e\sim 50-60$.  From Eqs.\ (\ref{phi_end})
and (\ref{N_e}), it is clear that, when $f_\phi\gg M_*$, our horizon
scale is affected only by the inflaton dynamics with $\phi\ll f_\phi$.
In this case, in particular, the inflaton amplitude for the COBE scale
$\phi_{\rm COBE}$ becomes much smaller than $f_\phi$.  Then, we can
approximate the inflaton potential as
$V\simeq\frac{1}{2}m_\phi^2\phi^2$, and all the predictions are the
same as those from the chaotic inflation model with a parabolic
potential with mass parameter $m_{\phi}$.  If $f_\phi\sim O(M_*)$, on
the contrary, $\phi_{\rm COBE}\sim O(f_\phi)$ and $V(\phi)$ cannot be
approximated as above.  In this case, we may observe some peculiar
signal due to the natural inflation potential (\ref{V_phi}).

Quantum fluctuation of $\phi$ during the inflation becomes the origin
of the density fluctuation.  At the time of the horizon exit, the
power spectrum of the curvature perturbation is given by
\begin{eqnarray}
{\cal P}^{1/2}(k) = 
\left[ \frac{H^2}{2\pi|\dot{\phi}|} \right]_{k=aH}.
\end{eqnarray}
${\cal P}(k)$ is often approximated using the power index $n_{\rm S}$;
denoting the curvature perturbation around $k\sim\bar{k}$ as ${\cal
P}(k)\simeq{\cal P}(\bar{k})\times (k/\bar{k})^{n_{\rm
S}(\bar{k})}$, the index is given by \cite{PRep314-1}
\begin{eqnarray}
n_{\rm S} (k) = 1 - 6\epsilon (k) + 2\eta (k)
\simeq 1 - \frac{M_*^2}{2f_\phi^2}
\left[
\frac{1+\cos^2(\phi/2\sqrt{2}f_\phi)}
{1-\cos^2(\phi/2\sqrt{2}f_\phi)} \right]_{k=aH}.
\label{n_S}
\end{eqnarray}
Thus, the spectrum deviates from the scale-invariant one if $\phi_{\rm
COBE}\sim O(f_\phi)$, which is realized when $f_\phi\sim O(M_*)$.
Importantly, the spectrum index becomes smaller than 1, and the
fluctuation for smaller scale is more suppressed compared to the case
of the scale-invariant power spectrum.

Now, we show the results of our numerical calculation.  In our
calculation, we first follow the evolution of the inflaton.  From the
period of the inflation to the time when $V(\phi)$ is well
approximated by the parabolic potential, Eq.\ (\ref{eom_phi}) is
solved.  After that period, we follow Eqs.\ (\ref{drho_phi}) and
(\ref{drho_rad}) until the inflaton decays.  Once the
radiation-dominated universe is realized, we use the simple scaling
law (i.e., $a^3s={\rm const.}$, with $s$ being the entropy density) to
obtain the normalization of the comoving momentum $k$
\cite{KolbTurner}.

Following the evolution of the inflaton field during the inflation, we
also calculate the curvature perturbation ${\cal P}$ as a function of
the comoving momentum $k$ as well as the spectrum index $n_{\rm S}$.
We compared ${\cal P}(k)$ from our numerical calculation with that
with the power-law approximation ${\cal P}^{\rm (approx)}(k)=
(k/k_{\rm COBE})^{n_{\rm S}(k_{\rm COBE})} {\cal P}(k_{\rm COBE})$,
and found that the power-law approximation is in a good agreement with
the numerical result.

The temperature fluctuation observed by COBE sets a constraint on the
primordial curvature perturbation ${\cal P}(k)$.  In order to discuss
the perturbation for the COBE scale, it is convenient to use the
parameter $\delta_H$, density perturbation at the time of the reentry
to the horizon.  For the matter dominated era (with sizable
contribution from the cosmological constant), $\delta_H$ is given by
\cite{ARAA30-499}
\begin{eqnarray}
\delta_H (k) = \frac{2}{5} \frac{g(\Omega_0)}{\Omega_0}
{\cal P}^{1/2} (k),
\label{d_H}
\end{eqnarray}
where the above expression is for the flat universe, and
\begin{eqnarray}
g(\Omega_0) = \frac{5}{2}\Omega_0
\left( \frac{1}{70} + \frac{209\Omega_0}{140}
- \frac{\Omega_0^2}{140} + \Omega_0^{4/7} \right)^{-1}.
\end{eqnarray} 
$\delta_H$ becomes the source of the temperature perturbation and is
constrained so that the COBE observations \cite{apj464-l1} are
reproduced \cite{APJ480-6}
\begin{eqnarray} 
|\delta_H(k_{\rm COBE})| &=& (1.94\pm 0.15) \times 10^{-5}
\nonumber \\ &&
\times \Omega_0^{-0.785-0.05\ln\Omega_0}
\exp \left[ -0.95(n_{\rm S}-1)-0.169(n_{\rm S}-1)^2 \right].
\label{d_H(COBE)}
\end{eqnarray}
This constrains the $f_\phi$ vs.\ $\Lambda$ plane.  For large
$f_\phi$, as we mentioned, all the physical quantities are determined
by the inflaton mass $m_\phi$.  In this case, the COBE scale exits the
horizon when $\phi\simeq 15M_*$, and ${\cal P}^{1/2}(k_{\rm
COBE})\simeq 7m_\phi/M_*$.  Using Eqs.\ (\ref{m_phi}), (\ref{d_H}),
and (\ref{d_H(COBE)}), the best-fit value of $\Lambda$ for large
enough $f_\phi$ is given by $\Lambda\simeq 8\times 10^{15}\ {\rm
GeV}\times (f_\phi/M_*)^{1/2}$ for $\Omega_0=0.4$.  When $f_\phi$
becomes small, higher order terms in the potential become more
important and the best-fit value of $\Lambda$ becomes smaller than the
above simple expression.

\begin{figure}[t]
\centerline{\epsfxsize=0.75\textwidth\epsfbox{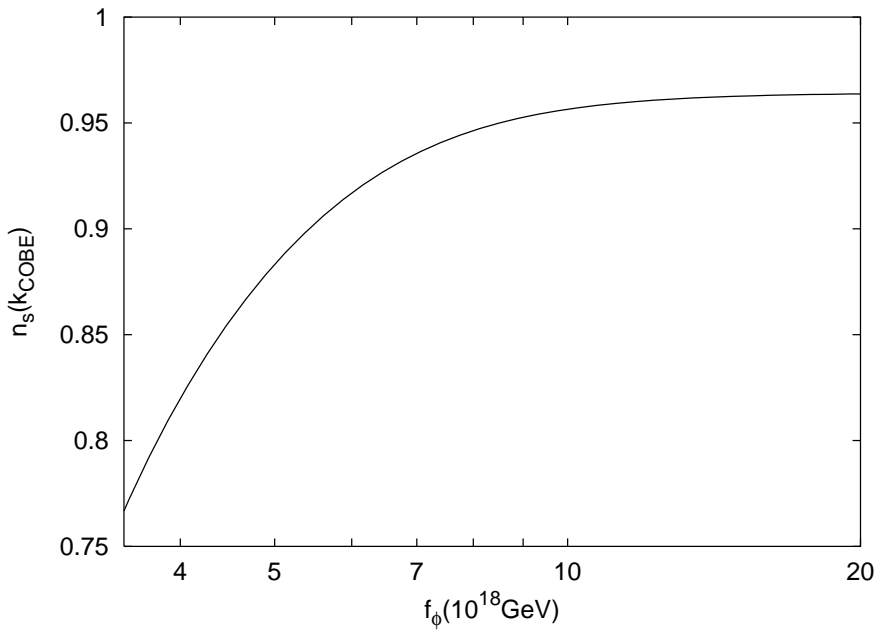}}
\caption{$n_{\rm S} (k_{\rm COBE})$ as a function of $f_\phi$.  The best-fit
value of $\Lambda$ for the COBE-scale normalization is used.}
\label{fig:index}
\end{figure}

COBE also sets a bound on the spectrum index.  The index for the COBE
scale is constrained as \cite{PRep314-1}
\begin{eqnarray}
n_{\rm S} (k_{\rm COBE}) = 1 \pm 0.2.
\label{n=1}
\end{eqnarray}
In Fig.\ \ref{fig:index}, we plot $n_{\rm S}(k_{\rm COBE})$ as a
function of $f_\phi$.  Here, for each $f_\phi$, we used the best-fit
value of $\Lambda$.\footnote{As indicated in Eq.\ (\ref{n_S}), $n_{\rm
S} (k_{\rm COBE})$ is insensitive to $\Lambda$ for fixed value of
$f_\phi$.  We numerically checked the validity of this statement.}
Using the constraint (\ref{n=1}), we can see that the natural
inflation with $f_\phi\lesssim 3.8\times 10^{18}\ {\rm GeV}$ is
inconsistent with the COBE observation.

Now, we are at a point to discuss the CMBR anisotropy for the $l$-th
multipole $C_l$, which is defined as \cite{HuThesis}
\begin{eqnarray}
\left\langle 
\Delta T(\vec{x},\vec{\gamma})
\Delta T(\vec{x},\vec{\gamma}')
\right\rangle =
\frac{1}{4\pi} \sum_l (2l+1) C_l P_l 
(\vec{\gamma} \cdot \vec{\gamma}'),
\end{eqnarray}
where $\Delta T(\vec{x},\vec{\gamma})$ is the temperature fluctuation
of the CMBR pointing to the direction $\vec{\gamma}$, and the average
is over the position $\vec{x}$.  Theoretically, $C_l$ is calculated
once the transfer function $T_l(k)$ is known:
\begin{eqnarray}
C_l \equiv \frac{2\pi}{l(l+1)} \tilde{C}_l
= \frac{4\pi}{(2l+1)} \int \frac{dk}{k} T_l^2 (k) {\cal P}(k).
\end{eqnarray}
We used the CMBfast package \cite{CMBfast} to calculate the transfer
function and obtained $\tilde{C}_l$ using the curvature perturbation
${\cal P}$ from the numerical calculation.  The cosmological
parameters used in our calculations are listed in Table
\ref{table:params}.  In particular, suggested from the recent studies
of the cosmological constant \cite{aph9904051}, we consider a model of
flat universe with non-vanishing cosmological constant.

\begin{table}[t]
\begin{center}
\begin{tabular}{lccc}
\hline\hline
{} & {$h$ \cite{aph0007187}} 
& {$\Omega_bh^2$ \cite{aph0007187}} 
& {$\Omega_0$ \cite{aph9904051}}\\
\hline
{Center value} & {0.65} & {0.019} & {0.4} \\
{Error (1-$\sigma$)} & {0.08} & {0.002} & {0.1} \\
\hline\hline
\end{tabular}
\caption{Cosmological parameters used in our calculation.
Here $\Omega_b$ and $\Omega_0$ are density parameters of baryon 
and total matter, respectively. We consider flat universe with cosmological 
constant.}
\label{table:params}
\end{center}
\end{table}

\begin{figure}[t]
\centerline{\epsfxsize=0.75\textwidth\epsfbox{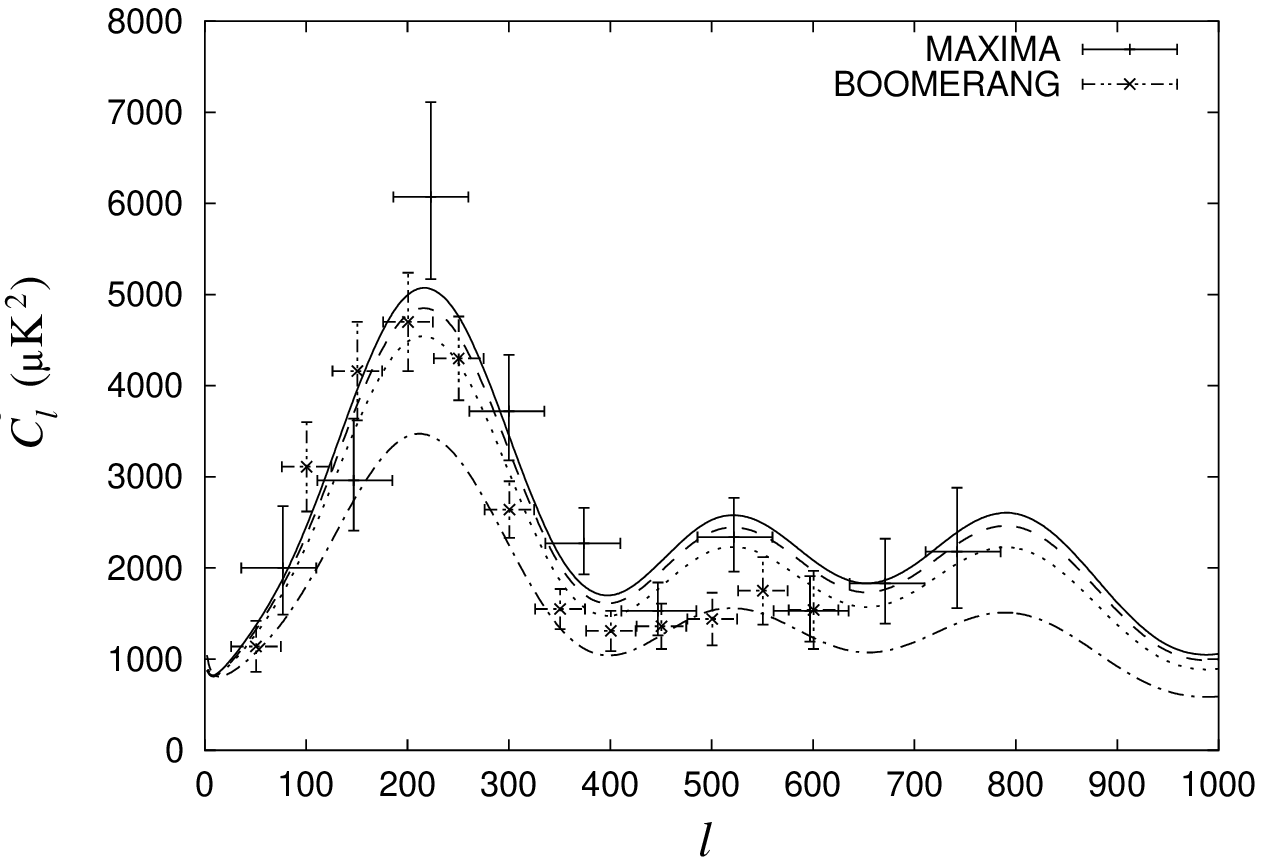}}
\caption{CMBR anisotropy for the $l$-th multipole.
The vertical axis is $\tilde{C}_l\equiv [l(l+1)/2\pi]C_l$.  Here, we
take $f_\phi= 10\times 10^{18}$\ GeV (solid), $8\times 10^{18}$\ GeV
(dashed), $6\times 10^{18}$\ GeV (dotted), and $4\times 10^{18}$\ GeV
(dot-dashed), and the best-fit values of $\Lambda$ for the COBE-scale
normalization are used.}
\label{fig:C_l}
\end{figure}

In Fig.\ \ref{fig:C_l}, we plot $\tilde{C}_l$ for several values of
$f_\phi$ with the best-fit value of $\Lambda$ for the COBE-scale
normalization.  With such a choice of $\Lambda$, $\tilde{C}_l$ for
smaller $l$ is almost independent of $f_\phi$.  For larger $l$,
however, $\tilde{C}_l$ is more suppressed for smaller $f_\phi$.  This
is because the curvature perturbation for smaller $f_\phi$ has a
smaller index parameter $n_{\rm S}$.  Notice that, with the best-fit
value of $\Lambda$ for the COBE-scale normalization, $\tilde{C}_l$
becomes almost independent of $f_\phi$ for $f_\phi\gtrsim 10^{19}\ 
{\rm GeV}$.

\begin{table}
\begin{center}
\begin{tabular}{cccc}
\hline\hline
{$l_{\rm eff}$} & {$\tilde{C}_l^{\rm obs} ({\rm \mu K}^2)$} & 
{$\sigma_{l,{\rm obs}} ({\rm \mu K}^2)$} & {Reference}
\\
\hline
{$3.1$}  & {$784$}   & {$473.5$} & {COBE} \\
{$4.1$}  & {$1156$}  & {$440.9$} & {COBE} \\
{$5.6$}  & {$630$}   & {$291$}   & {COBE} \\
{$8.0$}  & {$864.4$} & {$224.5$} & {COBE} \\
{$10.9$} & {$767.3$} & {$230.2$} & {COBE} \\
{$14.3$} & {$681.2$} & {$246.7$} & {COBE} \\
{$19.4$} & {$1089$}  & {$326$}   & {COBE} \\
{$51$}   & {$1140$}  & {$280$}   & {BOOMERANG} \\
{$101$}  & {$3110$}  & {$490$}   & {BOOMERANG} \\
{$151$}  & {$4160$}  & {$540$}   & {BOOMERANG} \\
{$201$}  & {$4700$}  & {$540$}   & {BOOMERANG} \\
{$251$}  & {$4300$}  & {$460$}   & {BOOMERANG} \\
{$301$}  & {$2640$}  & {$310$}   & {BOOMERANG} \\
{$351$}  & {$1550$}  & {$220$}   & {BOOMERANG} \\
{$401$}  & {$1310$}  & {$220$}   & {BOOMERANG} \\
{$451$}  & {$1360$}  & {$250$}   & {BOOMERANG} \\
{$501$}  & {$1440$}  & {$290$}   & {BOOMERANG} \\
{$551$}  & {$1750$}  & {$370$}   & {BOOMERANG} \\
{$601$}  & {$1540$}  & {$430$}   & {BOOMERANG} \\
{$77$}   & {$2000$}  & {$595$}   & {MAXIMA} \\
{$147$}  & {$2960$}  & {$615$}   & {MAXIMA} \\
{$223$}  & {$6070$}  & {$970$}   & {MAXIMA} \\
{$300$}  & {$3720$}  & {$580$}   & {MAXIMA} \\
{$374$}  & {$2270$}  & {$365$}   & {MAXIMA} \\
{$447$}  & {$1530$}  & {$290$}   & {MAXIMA} \\
{$522$}  & {$2340$}  & {$405$}   & {MAXIMA} \\
{$597$}  & {$1530$}  & {$360$}   & {MAXIMA} \\
{$671$}  & {$1830$}  & {$465$}   & {MAXIMA} \\
{$742$}  & {$2180$}  & {$660$}   & {MAXIMA}\\
\hline\hline
\end{tabular}
\caption{$C_{l}^{\rm obs}$ and $\sigma_{l,{\rm obs}}$ used in our
analysis, which are from COBE \cite{aph9702019}, BOOMERANG
\cite{Nature404-955} and MAXIMA \cite{aph0005123} experiments.}
\label{table:Cl}
\end{center}
\end{table}

Using the observations of the CMBR anisotropy by COBE, BOOMERANG and
MAXIMA, we can constrain the fundamental parameters $f_\phi$ and
$\Lambda$.  For this purpose, we calculate $\chi^2$ as\footnote{In our
calculation of $\chi^2$, we do not take account of the correlations
among data points.  Using the correlation matrices given by BOOMERANG
and MAXIMA, we derived a constraint on the $f_\phi$ vs.\ $\Lambda$
plane, and checked that the changes of the upper and lower bounds on
$\Lambda$ for fixed $f_\phi$ are less than a few \%.  We also neglect
the calibration uncertainties of the BOOMERANG and MAXIMA data sets
(20 \% for BOOMERANG and 8 \% for MAXIMA, 1-$\sigma$ in $C_l$
\cite{aph0007333}).  We found that the inclusion of the calibration
uncertainties may change the upper and lower bounds on $\Lambda$ by a
few \%.}
\begin{eqnarray}
\chi^2 (f_\phi, \Lambda) = 
\sum_{l}
\frac{[\tilde{C}_{l}^{\rm obs} - 
\tilde{C}_{l}^{\rm th}(f_\phi, \Lambda)]^2}
{\sigma_{l,{\rm obs}}^2 + \sigma_{l,{\rm th}}^2(f_\phi, \Lambda)}.
\end{eqnarray}
Here, $\tilde{C}_{l}^{\rm obs}$ is the observational data from COBE,
BOOMERANG and MAXIMA for the $l$-th multipole and $\sigma_{l,{\rm
obs}}$ is its error.  The sum is over $l$ available from the three
experiments. The data used in our analysis are listed in Table\ 
\ref{table:Cl}.  In addition, $\tilde{C}_{l}^{\rm th}(f_\phi,\Lambda)$
is the theoretical prediction to the $l$-th multipole as a function of
the fundamental parameters $f_\phi$ and $\Lambda$.  Furthermore, we
take account of the uncertainties in the cosmological parameters $h$,
$\Omega_bh^2$, and $\Omega_0$.  We varied these cosmological
parameters within the 1-$\sigma$ error and calculated the variations
in $\tilde{C}_{l}^{\rm th}(f_\phi,\Lambda)$.  We identified them as
systematic errors and added them in quadrature to evaluate
$\sigma_{l,{\rm th}}^2$.

\begin{figure}
\centerline{\epsfxsize=0.5\textwidth\epsfbox{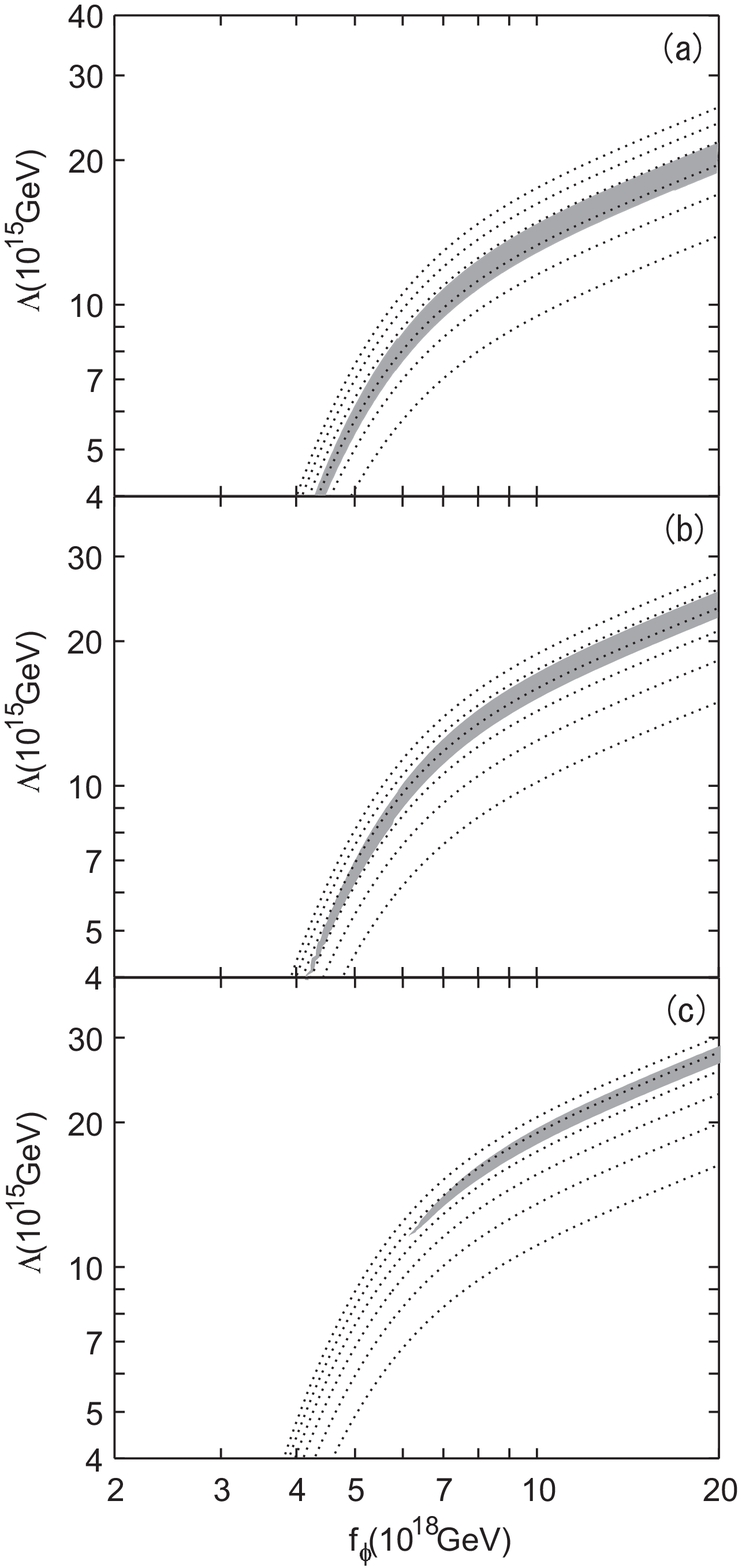}}
\caption{Constraint on the parameters $f_\phi$ and $\Lambda$.  The
shaded regions are for $\chi^2\leq 43$ for (a) $\tau=0$ (no
reionization), (b) $\tau=0.2$, and (c) $\tau=0.4$.  The dotted lines
are contours of constant $\sigma_8$  (0.4,
0.6, 0.8, 1.0, 1.2, and 1.4, from below).}
\label{fig:chi2}
\end{figure}

In Fig.\ \ref{fig:chi2}, we show the constraint on the $f_\phi$ vs.\ 
$\Lambda$ plane.  We shaded the region with $\chi^2\leq 43$ (which
corresponds to 95 \% C.L.\ allowed region for the
$\chi^{2}$-statistics with 29 degrees of freedom).  For a fixed
$f_\phi$, we obtain upper and lower bounds on $\Lambda$.  With
$f_\phi$ being fixed, the primordial curvature perturbation is an
increasing function of $\Lambda$, and hence $\Lambda$ is required to
have a relevant value to explain the observed size of the multipoles.
Notice that, for large enough $f_\phi$, the upper and lower bounds on
$\Lambda$ are proportional to $f_\phi^{1/2}$.  This is because the
observable quantities depend only on $m_\phi=\Lambda^2/\sqrt{2}f_\phi$
in this region.

Now, let us comment on the case with $f_\phi\gg M_*$.  In this case,
the index $n_{\rm S}$ becomes close to 1 and the theoretical
predictions on $\tilde{C}_l$ for $l\gtrsim 400$ become larger than the
observed ones.  However, the asymptotic value of $n_{\rm S}$ for
$f_\phi\gg M_*$ is 0.96, not exactly equal to 1.  Thus, the
constraints from the multipoles around the second peak is not as
severe as those for models with the scale-invariant Harrison-Zeldovich
spectrum.  In addition, since we used all the data points to evaluate
$\chi^2$ and required $\chi^2\leq 43$, the discrepancies for $l\gtrsim
400$ are not statistically significant enough to exclude the parameter
region with $f_\phi\gg M_*$.  Thus, in Fig.\ \ref{fig:chi2}, no upper
bound on $f_\phi$ is obtained.  However, it is interesting to consider
the constraint from the second peak.  As an example, following Ref.\ 
\cite{aph0008165}, we identified the highest data points for $l<400$
and $400\leq l\leq 600$ as the heights of the first and second peaks,
$\tilde{C}_l^{\rm 1st}$ and $\tilde{C}_l^{\rm 2nd}$, respectively.
Then, we combined the two data samples from BOOMERANG and MAXIMA to
obtain
\begin{eqnarray}
\tilde{C}_l^{\rm 2nd}/\tilde{C}_l^{\rm 1st} =  0.38 \pm 0.06.
\label{C(2nd/1st)}
\end{eqnarray}
Requiring $\tilde{C}_l^{\rm 2nd}/\tilde{C}_l^{\rm 1st}\leq 0.50$, we
obtain $f_\phi\lesssim 7\times 10^{18}\ {\rm GeV}$.  In the future,
more accurate measurements of the CMBR anisotropy will be able to
check the validity of this upper bound.

Another independent constraint is available from the cluster
abundance.  In Fig.\ \ref{fig:chi2}, we plotted the contours of the
constant $\sigma_8$, where $\sigma_8$ is the amplitude of mass density
fluctuations on the scale of $8h^{-1}\ {\rm Mpc}$.  The observed value
of $\sigma_8$ is given by $\sigma_8 = (0.56\pm 0.06)\Omega_0^{-0.47}$
\cite{aph9902245}.\footnote{Constraint on $\sigma_8$ is insensitive to
the index parameter $n_{\rm S}$ \cite{MNRAS262-1023}, and we neglect
its dependence on $n_{\rm S}$.} As one can see, in case without
reionization, $\sigma_8$ is consistent with the above-mentioned value
in the parameter region preferred by the CMBR anisotropy.

So far, we have not discussed the effect of the reionization after the
recombination.  Its effect is well parameterized by the following two
parameters: the optical depth $\tau$ and the red shift $z_{\rm ion}$
at the time of the reionization \cite{APJ479-568,aph9812125}.  Due to
the reionization, $\tilde{C}_l$'s with $l\gg z_{\rm ion}^{1/2}$ are
suppressed by the factor $e^{-2\tau}$ while those with small $l$ are
unchanged.  We calculated $\chi^2$ with the reionization with
$\tau=0.2$ and $0.4$.  For a fixed value of $\tau$, we took several
values of $z_{\rm ion}$ of $O(10)$, and checked that $\chi^2$ is
almost independent of $z_{\rm ion}$.  The results are shown in Fig.\ 
\ref{fig:chi2}.  Since the reionization effect reduces $\tilde{C}_l$,
larger value of the primordial perturbation is needed to obtain the
CMBR anisotropy consistent with the observations.  Thus, the preferred
value of $\Lambda$ becomes larger as $\tau$ increases, as shown in
Fig.\ \ref{fig:chi2}.  In addition, when $\tau=0$ and $n_{\rm S}=1$,
theoretical prediction for $\tilde{C}_l$ with $l\gtrsim 400$ becomes
larger than observations if we adopt the primordial curvature
perturbation suggested by COBE.  In this case, CMBR anisotropy prefers
a index $n_{\rm S}$ smaller than 1.  However, if the reionozatoin
effect is sufficient, it suppresses $\tilde{C}_l$ with large $l$ and
$n_{\rm S}<1$ makes the fit worse.  Thus, for large $\tau$, parameter
region with small $f_\phi$, where $n_{\rm S}$ becomes significantly
smaller than 1, is excluded.  In addition, consistency with the
constraint from the cluster abundance becomes worse for larger $\tau$,
as shown in Fig.\ \ref{fig:chi2}.

Finally, we briefly discuss possible improvement of the constraints
with future observations.  With MAP \cite{MAP} and PLANCK
\cite{PLANCK} experiments, much better observations of the CMBR
anisotropy will be obtained.  It has been pointed out that MAP and
PLANCK will determine the index $n_{\rm S}$ with $O(1\ \%)$ accuracy
\cite{MNRAS291-L33}.  In the natural inflation model, $n_{\rm S}$ is
sensitive to $f_\phi$.  For example, $5.2\times 10^{18}\ {\rm GeV}\leq
f_\phi\leq 5.7\times 10^{18}\ {\rm GeV}$ gives $0.89\leq n_{\rm S}\leq
0.91$.  In addition, $C_l$ itself will be determined much more
accurately, and hence the theoretical prediction on the CMBR
anisotropy will be more directly compared with the observation.  Thus
MAP and PLANCK will provide much better constraint on the natural
inflation model than the present one.

{\sl Acknowledgment:} One of the authors (TM) would like to thank T.\ 
Asaka and Y.\ Nomura for useful conversations.  This work is supported
by the Grant-in-Aid for Scientific Research from the Ministry of
Education, Science, Sports, and Culture of Japan (No.\ 12047201).


\begin{thebibliography}{100}

\bibitem{PRD23-347}
    A.H.\ Guth, Phys.\ Rev.\ {\bf D23} (1981) 347.

\bibitem{PLB108-389}
    A.D.\ Linde,
    Phys.\ Lett.\ {\bf B108} (1982) 389.

\bibitem{PRL48-1220}
    A.\ Albrecht and P.J.\ Steinhardt,
    Phys.\ Rev.\ Lett.\ {\bf 48} (1982) 1220.

\bibitem{PRL65-3233}
    K.\ Freese, J.A.\ Frieman and A.V.\ Olinto,
    Phys.\ Rev.\ Lett.\ {\bf 65} (1990) 3233.

\bibitem{PRD47-426}
    F.C.\ Adams, J.R.\ Bond, K.\ Freese, J.A.\ Frieman 
    and A.V.\ Olinto,
    Phys.\ Rev.\ {\bf D47} (1993) 426.

\bibitem{Nature404-955}
    P.\ de Bernardis et al.,
    Nature {\bf 404} (2000) 955.

\bibitem{aph0005123}
    S. Hanany et al.,
    astro-ph/0005123.

\bibitem{aph0007375}
    W.H.\ Kinney, A.\ Melchiorri and A.\ Riotto,
    astro-ph/0007375.

\bibitem{aph0008165}
    L.\ Covi and D.\ Lyth,
    astro-ph/0008165.

\bibitem{PRep314-1}
    See, for example, D.H.\ Lyth and A.\ Riotto,
    Phys.\ Rept.\ {\bf 314} (1999) 1.

\bibitem{KolbTurner}
    See, for example, E.W.\ Kolb and M.S.\ Turner,
    ``The Early Universe'' (Addison-Wesley, 1990).

\bibitem{ARAA30-499}
    S.M.\ Carroll, W.\ Press and E.L.\ Turner,
    Annu.\ Rev.\ Astron.\ Astrophys.\ {\bf 30} (1992) 499.

\bibitem{apj464-l1}
    C.L.\ Bennett et al., 
    Astrophys.\ J.\ {\bf 464} (1996) L1.

\bibitem{APJ480-6}
    E.F.\ Bunn and M.\ White,
    Astrophys.\ J.\ {\bf 480} (1997) 6.

\bibitem{HuThesis}
    See, for example, W.\ Hu,
    Ph.\ D Thesis (astro-ph/9508126).

\bibitem{CMBfast}
    U.\ Seljak and M.\ Zaldarriaga, 
    ``CMBFAST: A Microwave Anisotropy Code''
    (http://www.sns.ias.edu/$^\sim$matiasz/CMBFAST/cmbfast.html).

\bibitem{aph0007187}
    J.R.\ Primack,
    astro-ph/0007187.

\bibitem{aph9904051}
    M.S.\ Turner,
    astro-ph/9904051.

\bibitem{aph9702019}
    M.\ Tegmark and A.J.S.\ Hamilton,
    astro-ph/9702019.

\bibitem{aph0007333}
    A.H.\ Jaffe et al.,
    astro-ph/0007333.

\bibitem{aph9902245}
    P.T.P.\ Viana and A.R.\ Liddle,
    astro-ph/9902245.

\bibitem{MNRAS262-1023}
    S.D.\ White, G.\ Efstathiou and C.S.\ Frenk,
    Mon.\ Not.\ Roy.\ Astron.\ Soc.\ {\bf 262} (1993) 1023.

\bibitem{APJ479-568}
    W.\ Hu and M.\ White,
    Astrophys.\ J.\ {\bf 479} (1997) 568.

\bibitem{aph9812125}
    L.M.\ Griffiths, D.\ Barbosa and A.R.\ Liddle,
    astro-ph/9812125.

\bibitem{MAP}
    MAP home page (http://map.gsfc.nasa.gov/).

\bibitem{PLANCK} 
    PLANCK home page
    (http://astro.estec.esa.nl/SA-general/Projects/Planck).

\bibitem{MNRAS291-L33}
    J.R.\ Bond, G.\ Efstathiou and M.\ Tegmark,
    Mon.\ Not.\ Roy.\ Astron.\ Soc.\ {\bf 291} (1997) L33.

\end{thebibliography}
\end{document}